\RequirePackage{etex}
\documentclass{amsart}
\usepackage{amsmath, amsthm, amssymb,  mathrsfs, epsfig}
\usepackage{dcpic, pinlabel}

\usepackage{pinlabel}
\usepackage{MnSymbol}

\begin{document}

\newtheorem{theorem}{Theorem}[section]
\theoremstyle{definition}
\newtheorem{definition}[theorem]{Definition}

\title{A Topological Phase in a Quantum Gravity Model\\{\tiny{Solvay conference talk, October 2008}}}

\author{Michael H. Freedman}

\address{Microsoft Station Q, CNSI Bldg Rm 2243, University of California,\\
Santa Barbara, CA 93106-6105, U.S.A.\\
$^*$E-mail: michaelf@microsoft.com\\
}

%\bodymatter
\maketitle

\section*{\bf{Introduction}}

The concept of a topological phase may be traced to the interpretation [\cite{Thouless}] of the integer quantum Hall effect in terms of (what topologists call) a Chern class (or $K$-theory class) over the Brillion zone (momentum torus) and to Wilczek's realization [\cite{Wilczek}] that in $(2+1)$-dimensions, anyon statistics was a possibility to be considered on an equal footing with the more familiar fermionic and bosonic statistics.

The discovery of the fractional quantum Hall effect, Laughlin's wave function, charge fractionalization, and Halperin's realization [\cite{Halperin}] that the whole package must necessarily include anyonic statiatics, had by the mid 1980's presented us with a rather firm ``existence proof'' of topological phases.  Unlike in mathematics, ``proof'' preceded definition.

I would claim that two decades later we do not have a suitably general definition of what a topological phase is, or more importantly, any robust understanding of how to enter one even in the world of mathematical models.  The latter is, of course, the more important issue and the main subject of this note.  But a good definition can sharpen our thinking and a poor definition can misdirect us.  I will not attempt a final answer here but merely comment on the strengths and weaknesses of possible definitions and argue for some flexibility. In particular, I describe a rather simple class of ``quantum gravity'' models which are neither lattice nor field theoretic but appear to contain strong candidates for topological phases.

What is a topological phase?  The easiest answer is that a topological phase is a system whose effective low energy theory is governed by a Chern-Simons Lagrangian.  This answer is extremely efficient but limiting as it overlooks Dijkgraaf-Witten finite group [\cite{F}] TQFTs which can be very interesting even in the absence of a Chern-Simons term (in this case a twist class $\beta \in H^4(BF;U(1))$) and possibly other, as yet unknown, topological structures.  This definition would be a bit like defining a group to be a set of matrices with certain properties; the definition is too limiting since there are non-matrix groups.

Similarly, any definition which contains phrases such as ``spin-charge separation,'' ``fractional charge,'' ``point-like excitations,'' and ``string-operator'' presume too much: that electrons carry the relevant microscopic degrees of freedom or that the system is quasi $(2+1)$-dimensional, and may even unnecessarily exclude novel states of electrons confined in two dimensions.  I prefer a spectral definition (but will also criticize it!).
\begin{definition} \label{defnA}
Let $\mathcal{H}$ be a Hilbert space with local degrees of freedom.  A Hamiltonian $H: \mathcal{H} \to \mathcal{H}$ is said to describe a {\em topologial phase} if:
\begin{enumerate}
\item $H$ has degenerate ground state
\item $H$ has a gap to the first excited state
\item (1) and (2) are ``stable'' with respect to any sum of local perturbations.
\end{enumerate}
\end{definition}

To explain: ``gap'' refers to a constant size energy gap in the thermodynamic limit and ``stable'' means, chiefly, that the splitting of the ground state degeneracy is exponentially small in a length scale (and also excludes the unlikely possibility that the gap closes instantly, in the thermodynamic limit, under perturbation).  Equivalent to condition (3) is a more ``cryptographic'' condition (3'): for any local operator $\mathcal{O}$, the composition $G \overset{inc}{\longrightarrow} \mathcal{H} \overset{\mathcal{O}}{\longrightarrow} \mathcal{H} \overset{inc^\dagger}{\longrightarrow} G$ is a scalar (or exponentially close to one).  What I like about this definition is that is agnosist as to the type of local degrees of freedom, the dimension, and the nature and shape (e.g. could be string-like) of excitations.  Also attractive is that equivalence classes of phases may simply be defined as the deformation classes of $H$ subject to (1), (2), and (3).

There are two things I do {\em not} like.  First, to achieve a ground state degeneracy, periodic boundary conditions (e.g. wrap the quantum medium up to a closed surface) must be invoked.  Since a topological phase is a {\em local} concept it is disconcerting to need a {\em global} ingredient in its definition.  This state of affairs is like having a definition of hyperbolic geometry that did not work locally but only made sense for closed surfaces.  (To keep the definition local, one might try using the existence of the constant $-\log \mathcal{D}$ term [\cite{LW1,KP}] in the von Neumann entropy of partial trace(ground state $\psi_0$), $S(P_A)$, but this approach has not yet been adequately explored.)

Second, in the example, the ``quantum gravity'' Hamiltonian $H_{qg}$ which I will now explain, will have gapless ``gravity waves'' which appear to have no interaction with the gapped topological degrees of freedom.  Nonetheless, I would like to consider $H_{qg}$ as defining a topological phase.  Indeed, it appears to be the simplest route to realizing Turaev-Viro TQFTs (previously described in [\cite{LW2}] using 12-body interactions).  Thus, while definition \ref{defnA} heads in the right direction, it is still too restrictive.

\section*{\bf{$H_{qg}$ Described for the $Dfib$ Phase}}

$H_{qg} = H_{qg}^0 + \lambda V$ should be thought of as a bundle of Hamiltonians over the moduli space of metrics on a surface $\Sigma$ (say a torus).  The terms of $H_{qg}^0$ are fusion constraints acting within fibers and $F$-moves which act between fibers.  The Levin-Wen [\cite{LW2}] 12-body plaquet term required to define the phase arises at second order from a perturbation $\lambda V$ which virtually excites an electric pair $(\tau \otimes 1, \tau \otimes 1)$ or $(1 \otimes \tau, 1 \otimes \tau)$ in the notation of [\cite{F}].

\begin{figure}[]
\begin{center}
\epsfig{file=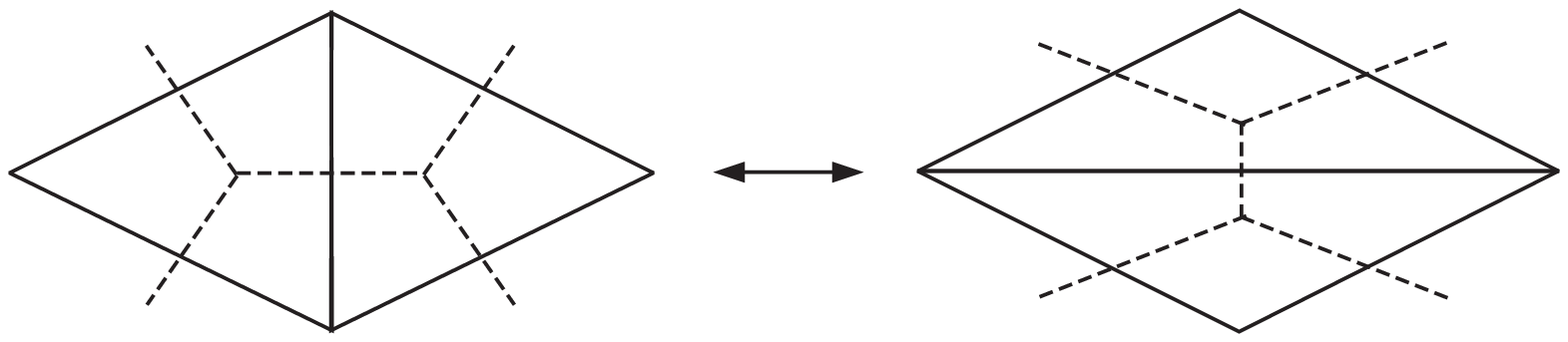,width=3in}
\end{center}
\caption{}
\label{move}
\end{figure}

We step back and specify the Hilbert space $\mathcal{H}$.  $\mathcal{H}$ is spanned by kets which are pairs $|(\Delta, S)\rangle$, where $\Delta$ is a triangulation of $n$ (fixed) triangles of the surface $\Sigma$ and $S$ is a labeling by particle types, in this case from the set $\{1,\tau\}$, of the dual net $N$ to $\Delta$.  We consider two triangulations $\Delta$ and $\Delta'$ (and their dual nets $N$ and $N'$) equivalent if they are isotopic on $\Sigma$ (i.e. we can slide one onto the other).  The dynamics on the set $\mathcal{N}_n$ of nets dual with $n$-vertex triangulations consist of the move in figure \ref{move} and is known to mix algebraically $\lambda_1 (\mathcal{N}_n) :: \frac{1}{n^\gamma}$, $\gamma$ positive.

\noindent Here, $\mathcal{N}_n$ is regarded as an abstract graph with vertices $\Delta_n$ and edges given by moves.  $\lambda_1$ is the first eigenvalue of the graph Laplacian.  To repeat, a ket is a net $N$ with some edges marked 1 and some edges marked $\tau$.  $N$ may be a regular honeycomb $N^0$, or a quite irregular $N^i$.

Next we define $H_{qg}^0$.  First, it enforces fusion rule terms at each vertex of each $N^i$ by penalizing the illegal Fibonacci fusion (and its symmetries ${}^{1}_{\tau}\rightY \,${\scriptsize{1}}).  Second, it contains terms between states of adjacent nets $N$ and $N'$ which enforce the unitary $F$-symbol $\displaystyle \left| \begin{array}{lr} \tau^{-1} & \tau^{1/2} \\ \tau^{1/2} & -\tau^{-1} \end{array} \right|$, $\tau = \frac{1 + \sqrt{5}}{2}$.

\begin{figure}[]
\labellist \normalsize\hair 2pt

  \pinlabel $v=$ at 15 220
  \pinlabel $w=$ at 15 75
  \pinlabel $-\tau^{-1}$ at 215 220
  \pinlabel $-\tau^{1/2}$ at 370 220
  \pinlabel $-\tau^{1/2}$ at 215 75
  \pinlabel $+\tau^{-1}$ at 370 75

\endlabellist
\centering
\includegraphics[scale=0.325]{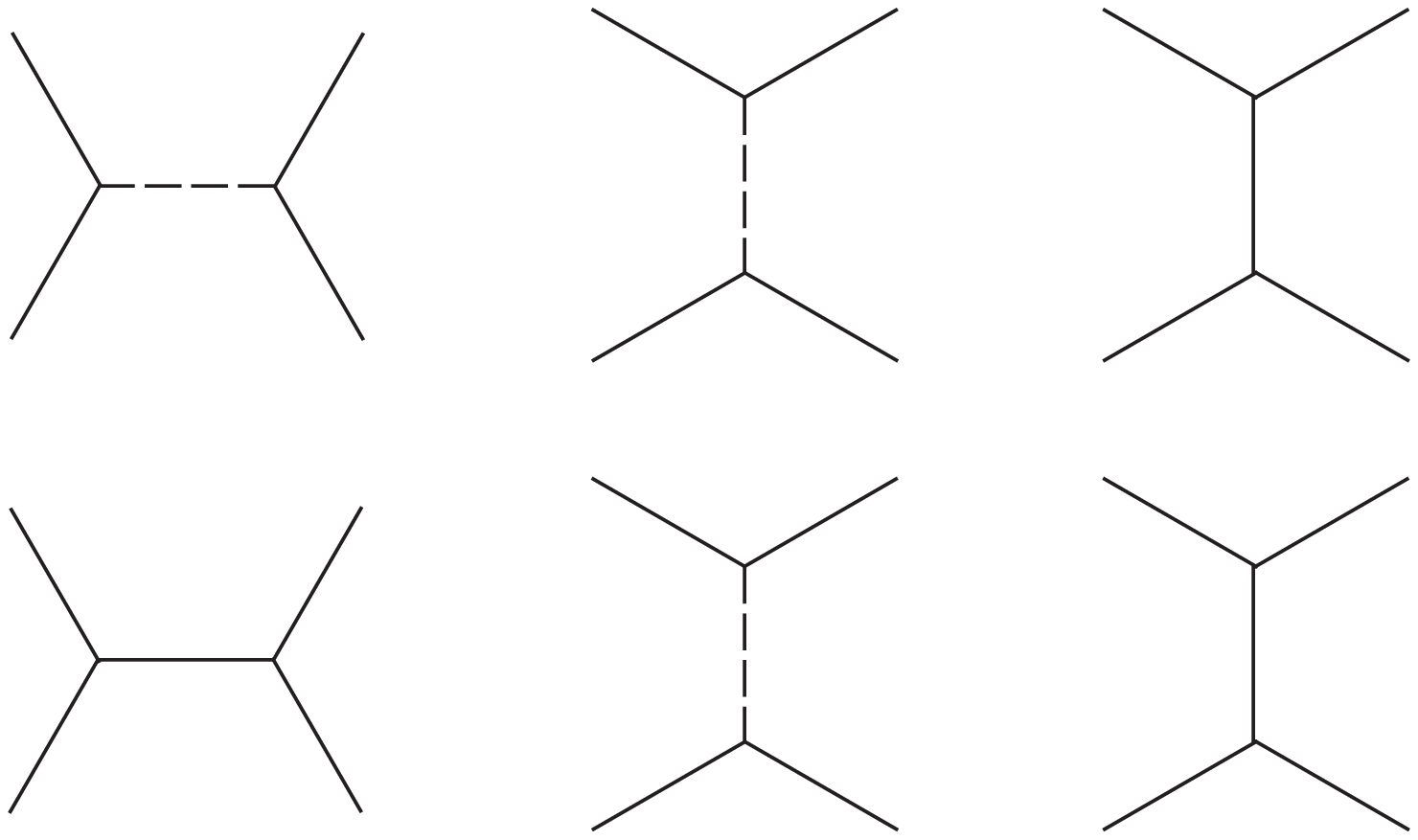}
\caption{} \label{states}
\end{figure}

Let $v$, $w$ be the normalized states of $H$, shown in figure \ref{states}.  The second terms of $H_{qg}^0$ are of the form $(id - |v\rangle\langle v|)$ and $(id - |w\rangle\langle w|)$.  (In figure \ref{states}, solid lines carry the $\tau$ particle label and dotted lines the trivial label.)

Finally, the perturbation $\lambda V$ creates an ``electric pair'' (either $(\tau \otimes 1, \tau \otimes 1)$ or $(1 \otimes \tau, 1 \otimes \tau)$ by breaking a $\tau$-labeled string resulting in a pair of excitations (fusion rule violations) [\cite{F}]).  $H_{qg} = H_{qg}^0 + \lambda V$.

Let us discuss the spectrum of $H_{qg}^0$ first.  $H_{qg}^0$ is positive semi-definite and its ground state manifold consists of the states $\psi$ with $\langle \psi | H_{qg}^0 | \psi \rangle = 0$.  Such a wave function $\psi$ is completely determined by its restriction to a sample net $N_0$ via the $F$-symbols.  (Importantly, $\psi$ is not {\em over} determined (frustrated) since the $F$-symbol satisfies the pentagon equations.)  The ground state manifold may be classified according to the number of magnetic particles $\tau \otimes \tau$ (of which, in our example system, there is only one type).  Since we have only imposed fusion and $F$-moves there is no energy penalty for $\tau \otimes \tau$ charges, provided they are not frustrated and instead are allowed to roam ergodically according to the moves ($F$) which link adjacent nets.  The magnetic charges on $N^0$ can return arbitrarily permuted, so the only zero energy (unfrustrated) states with $j$-magnetic charges, $j \geq 2$, are the ones that have equal amplitude for all positions of the $j$ charges (on all $n$-vertex nets).  One may think of the $j$-magnetic charge states as dispersing into momentum bands, although this terminology is not precise since translation does not even make sense on the general $N_i$.  Nevertheless, it is true that for each $j \neq 1$, the $j$-magnetic charge states have four zero energy (``zero momentum'') representatives $\{\psi_j\}$ on the torus.  Above each of these are ``bands'' and gapless ``gravity waves.'' The latter are ``magnons'' or phase oscillation across the (not very tightly bound) graph $\mathcal{N}_n$.

\begin{figure}[]
\labellist \small\hair 2pt

  \pinlabel $\psi_0$ at 15 25
  \pinlabel $\psi_2$ at 120 25
  \pinlabel $\psi_3$ at 225 25
  \pinlabel $j=0$ at 50 10
  \pinlabel $j=2$ at 155 10
  \pinlabel $j=3$ at 258 10
  \pinlabel $\text{gravity}$ at 450 55
  \pinlabel $\text{waves}$ at 450 45
  \pinlabel $\text{pair in}$ at 200 55
  \pinlabel $\text{``band''}$ at 200 45
  \pinlabel $\text{triple in}$ at 310 55
  \pinlabel $\text{``band''}$ at 310 45

\endlabellist
\centering
\includegraphics[scale=0.75]{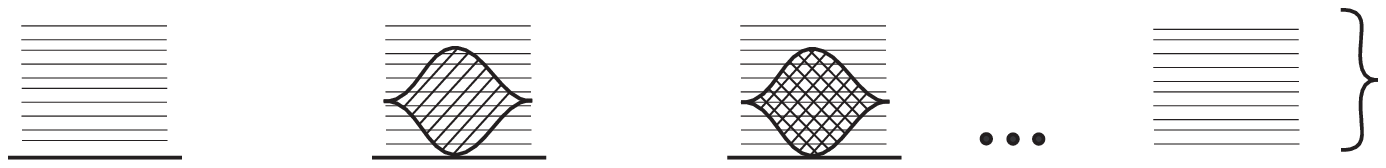}
\caption{} \label{gravwave}
\end{figure}

Now consider a perturbation $\lambda V$ which (virtually) pulls an electric pair (say $(\tau \otimes 1, \tau \otimes 1)$) out of the vacuum.  Because of the {\em nontrivial} mutual statistics between the magnetic ($\tau \otimes \tau$) and electric ($\tau \otimes 1$) excitations, a frustration arises which increases the cost of the electric pair $\psi_j^{e,e^*}$ in the presence of magnetic particles.  For small $j$ the effect is roughly linear: $$\langle \psi_j^{e,e^*} | H_{qg}^0 | \psi_j^{e,e^*} \rangle - \langle \psi_0^{e,e^*} | H_{qg}^0 | \psi_0^{e,e^*} \rangle \approx j \alpha$$ for some $\alpha > 0$ and where we have set $\langle \psi_0^{e,e^*} | H_{qg}^0 | \psi_0^{e,e^*} \rangle =1$.

In the presence of $\lambda V$ a second order virtual process will lower the energy of $\psi_j$ by $\sim \lambda^2(1+j\alpha)^{-1} \approx \lambda^2(1-j\alpha)$ producing an energy splitting separating the ``true vacuum'' $\psi_0$ from the ``magnetically charged'' vacuum by $\approx 2\lambda^2 \alpha$.  ($\alpha$ is the energy scale of the $F$-symbol constraint divided by the square of the minimal number of moves (16) required to move a plaquet around a closed loop and across an ``electric string'' in any family of nets.  This analysis is, so far, quite superficial.  We should also consider the corresponding energy reductions induced by second order virtual processes between energy $\delta$ gravity wave states $\psi_{0,\delta}$ and $\psi_{j,\delta}$ above their respective vacua and the corresponding electric excitations $\psi_{0,\delta}^{e,e^*}$ and $\psi_{j,\delta}^{e,e*}$.  However, the phase variations of any $\psi_{j,\delta}$, $j \geq 0$, over the number of moves (16) required to braid a $\tau \otimes \tau$ around electric strings can be made arbitrarily small by picking $\delta$ close to zero.  Thus, the preceding argument adapts to show that the gravity wave states over $\psi_0$ are reduced in energy by this process more than the corresponding states over $\psi_j$, $j \geq 2$.  (Details will appear in a joint paper with M. Troyer, whom I also thank for discussion on the concepts of this note.)  Thus, the perturbation picks out the sector containing the true vacuum $\psi_0$ as lowest energy.

A comparison of $H_{qg}$ to the exactly solved Levin-Wen Hamiltonian $H_{LW}$ is instructive.  The ground states (in the thermodynamic limit) are expected to be bijective.  The excitations of $H_{qg}$ are, in contrast to $H_{LW}$, mobile.  To build point-like, confined excitations ``wave packets'' will need to be formed.  Combinatorial recoupling arguments suggest that if such packets are confined in potential wells and braided, the L-W (i.e. Jones) braid representation should be exactly realized (in the thermodynamic limit).  Thus, we expect that the entire topological structure, the TQFT, represented by $H_{LW}$ is recaptured by $H_{qg}$.

$H_{qg}$ is not a ``lattice Hamiltonian.''  In particular, it is not defined on a ``tensor product'' Hilbert space (but rather a fiber-wise direct sum of these, one for each net in $\mathcal{N}_n$).  Thus, it is not precise to assert that $H_{qg}$ is ``$k$-body'' for any $k$, but it is evidently quite simple. One may say that the flux (plaquet) term of $H_{LW}$ (which is 12-body) has been simulated by more local interactions, but to achieve this we have resorted to a context where the lattice itself fluctuates and must be counted among the dynamic variables.  Hence the sobriquet: quantum gravity Hamiltonian.

\bibliographystyle{plain}
\bibliography{topphase}

\end{document}